\documentclass[preprint,amsmath,amssymb,aps]{revtex4-2}
\usepackage{graphicx}
\usepackage{dcolumn}
\usepackage{bm}
\usepackage[utf8]{inputenc}
\usepackage[T1]{fontenc}
\usepackage{mathptmx}
\usepackage{etoolbox}
\usepackage{braket}
\usepackage{soul}

\begin{document}
\title{An isotropic antenna based on Rydberg atoms}%
\author{Shaoxin Yuan}%
\thanks{These authors contributed equally to this work}
\author{Mingyong Jing} 
\thanks{These authors contributed equally to this work}
\email{jmy@sxu.edu.cn}
\altaffiliation[Also at ]{State Key Laboratory of Precision Measurement Technology and Instruments, Department of Precision Instrument, Tsinghua University, Haidian, Beijing 100084, China}

\author{Hao Zhang}

\author{Linjie Zhang}
\email{zlj@sxu.edu.cn}

\author{Liantuan Xiao}

\author{Suotang Jia}
\affiliation{State Key Laboratory of Quantum Optics and Quantum Optics Devices, Institute of Laser Spectroscopy, Shanxi University, Taiyuan, Shanxi 030006, China}
\affiliation{Collaborative Innovation Center of Extreme Optics, Shanxi University, Taiyuan, Shanxi 030006, China}

\date{\today}%

\begin{abstract}
Governed by the hairy ball theorem, classical antennas with isotropic responses to linearly polarized radio waves are unrealizable. This work shows that the antenna based on Rydberg atoms can theoretically achieve an ideal isotropic response to linearly polarized radio waves; that is, it has zero isotropic deviation. Experimental results of isotropic deviation within 5 dB, and 0.3 dB achievable after optimization, in microwave and terahertz wave measurements support the theory and are at least 15 dB improvement than the classical omnidirectional antenna. Combined with the SI traceable and ultrawideband property, the ideal isotropic response will make radio wave measurement based on atomic antenna much more accurate and reliable than the traditional method. This isotropic atomic antenna is an excellent example of what a tailored quantum sensor can realize, but a classical sensor cannot. It has crucial applications in fields such as radio wave electrometry.
\end{abstract}

\maketitle

\section{Introduction}
Metrology is the mother of science; the advancement of technology is inseparable from precision measurement. The precision measurement of many physical quantities has rapidly developed in recent years, and a typical example is the measurement of time/frequency \cite{Hall2006}. However, things go awry when measuring the radio frequency (RF) electric field (E-field). An ideal antenna with an isotropic reception pattern is crucial in the precision measurement/calibration of E-field  \cite{Hartansky2016,Holloway2006}; otherwise, even a mere change in how they are held can introduce significant measurement errors.

Unfortunately, isotropic antennas that can provide full spatial coverage are currently fictional and theoretically cannot be realized based on the classical metal antenna. This is because the transverse electric field in the far field region cannot be uniform over a sphere if the field is linearly polarized everywhere, or more basically, is governed by the hairy ball theorem: \textit{An even dimensional topological sphere $\mathbb{S}^n$ does not admit any continuous vector field of non-zero tangent vectors} \cite{Burns2005,Eisenberg1979,Milnor1978}. According to the antenna reciprocity theorem \cite{Stutzman2012}, the hairy ball theorem applies to both receiving and radiating antennas. Scientists have conducted many studies based on classical techniques to realize null-free quasi-isotropic antennas, including folded dipoles \cite{Liu2020,Kim2017,Ouyang2017}, magnetic dipoles \cite{Li2017}, orthogonal dipoles \cite{Kim2020,Su2020,Pan2017}, etc. However, they can only respond approximately isotropically, are complex in structure \cite{Yektakhah2017}, and are sensitive to wavelength \cite{Kim2017,Li2017}. In the classical case, the only way to reduce the error due to non-isotropic response is to calibrate the antenna; however, the calibration faces a causal dilemma \cite{Holloway2022,Simons2018}.

Here we introduce another alternative path: an isotropic antenna based on Rydberg atoms  \cite{Sedlacek2012,Jing2020,Ding2022}, which theoretically allows an ideal isotropic response to a linearly polarized field to be measured. The most significant difference between the classical antenna and the atomic antenna is that the classical antenna is made by conducted material, in which electrons are delocalized and can move or distribute freely in the three-dimensional Euclidean space to form a macroscopic charged topological 2-sphere $\mathbb{S}^2$. In contrast, the atomic antenna is constructed by many Rydberg atoms. The electron of Rydberg atoms is gravitated and localized around the atom; thus, it can only be regarded as a single point and cannot form a topological surface. Therefore, the hairy ball theorem does not apply to atomic antennas, making the theoretical isotropic response possible. The following sections will give more rigorous theoretical analysis and experimental verification. This ideal isotropic antenna has essential applications in electric field calibration \cite{Holloway2014,Jiao2017,Lim2023} and some modern communications where full spatial coverage and/or uniform signal reception is required \cite{Kim2020,Deng2013}.

\section{Intuitive Theory of Isotropic Atomic antenna}
We use an intuitive theory to illustrate the working principle of the isotropic atomic antenna. This intuitive theory only considers that the RF field couples two Rydberg energy levels and constructs the Hamiltonian of the light-atom interaction. The eigenanalysis of the interaction Hamiltonian shows that the eigenvalue of this Hamiltonian is unaffected by the polarization direction (equivalent, direction of propagation) of the RF field under the configuration that meets the requirements of the application scenario.
\begin{figure}
\includegraphics[width=0.9\columnwidth]{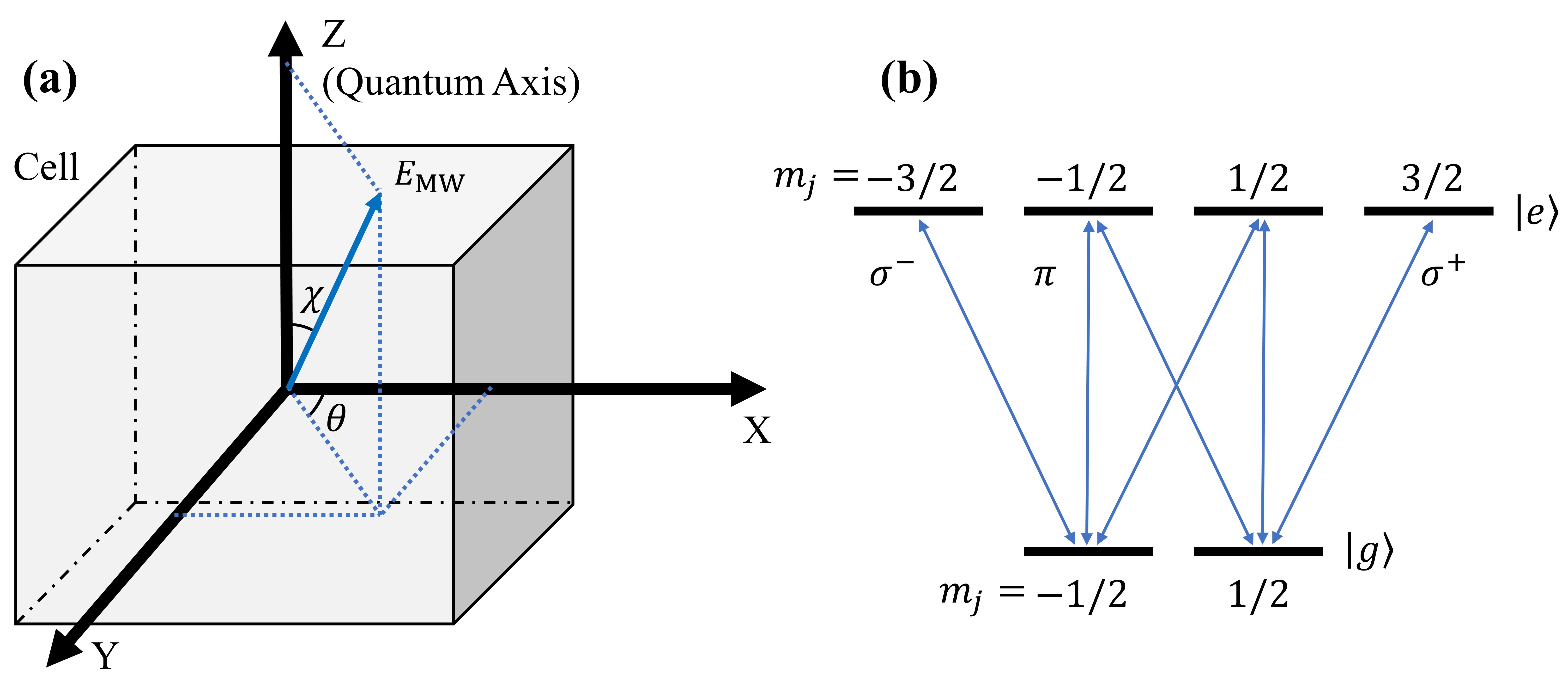}
\caption{\label{fig1}Schematic of intuitive theory. a). Schematic of RF E-field vector (polarization) orientation. The Z-axis is chosen as the quantum axis. $\chi$ (inclination) is the angle between the RF field polarization and the quantum axis. $\theta$ (azimuth) is the angle between the X-axis and the orthogonal projection of the RF field polarization on the XY plane. b). Level diagram of intuitive theory, $\ket{g}$ refers to the first Rydberg state with an angular momentum of $J=1/2$, and $\ket{e}$ refers to the second Rydberg state with an angular momentum of $J=3/2$. The RF fields couple each sublevel of $\ket{g}$ and $\ket{e}$ through $\sigma^\pm$ or $\pi$ transition.}
\end{figure}

As shown in Fig. \ref{fig1}(b), the two energy levels of the atom have angular momentum of $J=1/2$ and $J=3/2$, may correspond to actual Rydberg levels of $nS_{1/2}$ and $n'P_{3/2}$, respectively. The atomic state with $J=1/2$ allows two sub-energy levels with $m_j = -1/2$ and $1/2$, and the atomic state with $J=3/2$ allows four sub-energy levels with $m_j = -3/2$, $-1/2$, $1/2$, and $3/2$. According to the transition selection rules, the RF field couple these sub-levels through $\pi$ and $\sigma^\pm$ transitions with $\delta m_j=0$ and $\pm 1$. The strength of the individual transitions depends on the orientation of RF field polarization. Figure \ref{fig1}(a) shows the orientation of the RF field polarization vector. We chose Z-axis as the quantum axis for convenience, and the orientation of field polarization can be fully described by the inclination $\chi$ and azimuth $\theta$ in the spherical coordinate. According to this configuration, the field-atom interaction Hamiltonian in Cartesian coordinate has the following form:

\begin{eqnarray}
  H & = & \begin{bmatrix}
  M_g&M^\dagger_I \\
 M_I &M_e
\end{bmatrix},
\end{eqnarray}
where $M_g$ represents the Hamiltonian of bare $\ket{g}$ state, $M_e$ represents the Hamiltonian of bare $\ket{e}$ state and $M_I$ represents the Hamiltonian of the interaction. $M_g$, $M_e$ and $M_I$ take the form of:
\begin{eqnarray}
M_g & = &
\begin{bmatrix}
  0&0 \\
 0 &0
\end{bmatrix},
\end{eqnarray}

\begin{eqnarray}
M_e=
\begin{bmatrix}
 -\Delta  & 0 & 0 & 0\\
 0 &  -\Delta & 0 &0 \\
 0 & 0 &  -\Delta & 0\\
 0 & 0 & 0 & -\Delta
\end{bmatrix},
\end{eqnarray}

\begin{eqnarray}
&M_I = \frac{\Omega}{4}
\begin{bmatrix}
 -\sqrt{3} e^{i (\theta +\phi )} \sin\chi  & 0 \\
 2 \cos\chi  & -e^{i (\theta +\phi )} \sin\chi  \\
 e^{-i (\theta -\phi )} \sin\chi  & 2 \cos\chi  \\
 0 & \sqrt{3} e^{-i (\theta -\phi )} \sin\chi  \\
\end{bmatrix},
\end{eqnarray}
where $\Delta$ is the frequency detuning, $\sqrt{6}\Omega$ is the reduced Rabi frequency of the RF field, and $\phi$ is the relative phase angle between two field polarization components, one of which along the quantum axis and the other along the XY plane.

This Hamiltonian gives six eigenvalues ($\lambda_1$ to $\lambda_6$) of 
\begin{eqnarray}
\lambda_1&=&\lambda_2=-\Delta \nonumber\\
\lambda_3&=&-\frac{1}{2}(\Delta+\sqrt{\Delta ^2+\Omega^2 (1+\sin\chi \cos\chi \sin\phi)}) \nonumber\\
\lambda_4&=&-\frac{1}{2}(\Delta -\sqrt{\Delta^2+\Omega^2(1+\sin\chi \cos\chi \sin\phi)}).\\
\lambda_5&=&-\frac{1}{2}(\Delta+\sqrt{\Delta^2+\Omega^2(1-\sin\chi \cos\chi \sin\phi)}) \nonumber\\
\lambda_6&=&-\frac{1}{2}(\Delta -\sqrt{\Delta^2+\Omega^2 (1-\sin\chi \cos\chi \sin\phi)})\nonumber
\end{eqnarray}
Consider the specific application scenario: measuring the amplitude of a linear polarized RF field with known frequency, the $\Delta$ will be known, and the relative phase angle between different polarization components should be zero, i.e., $\phi=0$. Then the eigenvalues will reduce to
\begin{eqnarray}
\lambda_1&=\lambda_2&=-\Delta \nonumber\\
\lambda_3&=\lambda_5&=-\frac{1}{2}(\Delta+\sqrt{\Delta ^2+\Omega ^2}) ,\\
\lambda_4&=\lambda_6&=-\frac{1}{2}(\Delta-\sqrt{\Delta ^2+\Omega ^2})\nonumber
\end{eqnarray}
and the energy difference between $\lambda_{3(5)}$ and $\lambda_{4(6)}$ is 
\begin{equation}
\Delta_{\rm AT}=\sqrt{\Delta^2+\Omega^2},\label{eq.ats}
\end{equation}
where $\Delta_{\rm AT}$ can be direct measured by Autler-Townes (A-T) splitting spectrum in experiments \cite{Autler1955,Anisimov2011}. With known of $\Delta$ and $\Delta_{\rm AT}$, the following measurement equation can obtain the field amplitude:
\begin{equation}
E_{\rm MW}=\frac{\hbar}{\mu}\sqrt{\Delta_{\rm AT}^2-\Delta^2}.\label{eq.measurement}
\end{equation}
Equation \ref{eq.measurement} shows that $\chi$ and $\theta$ do not affect measured results; that is, the incident direction of the RF field does not affect the measured field amplitude by the atomic antenna and the atomic antenna has an isotropic response. Equation \ref{eq.measurement} also shows the SI-traceable feature unique to the atomic antenna \cite{Sedlacek2012}: Its measurement equation only involves frequency measurement and can be directly traced to the Planck's constant through the transition dipole moment, which can be accurately calculated through the Alkali.ne Rydberg Calculator (ARC) package  \cite{Sibalic2017}. Here we define the normalized receive-gain ($G_{\rm norm}$) of atomic antenna in decibels as
\begin{equation}
G_{\rm norm}=20\log \frac{\Delta_{\rm AT}/E_{\rm MW}}{{\rm max}(\Delta_{\rm AT}/{E_{\rm MW}})},
\end{equation}
where ${\rm max}(x)$ is the the largest value of set $x$. Theoretically, for the atomic antenna, the normalized gain is a constant of 0 dB.

\section{Experimental Results}
\begin{figure}
\includegraphics[width=0.9\columnwidth]{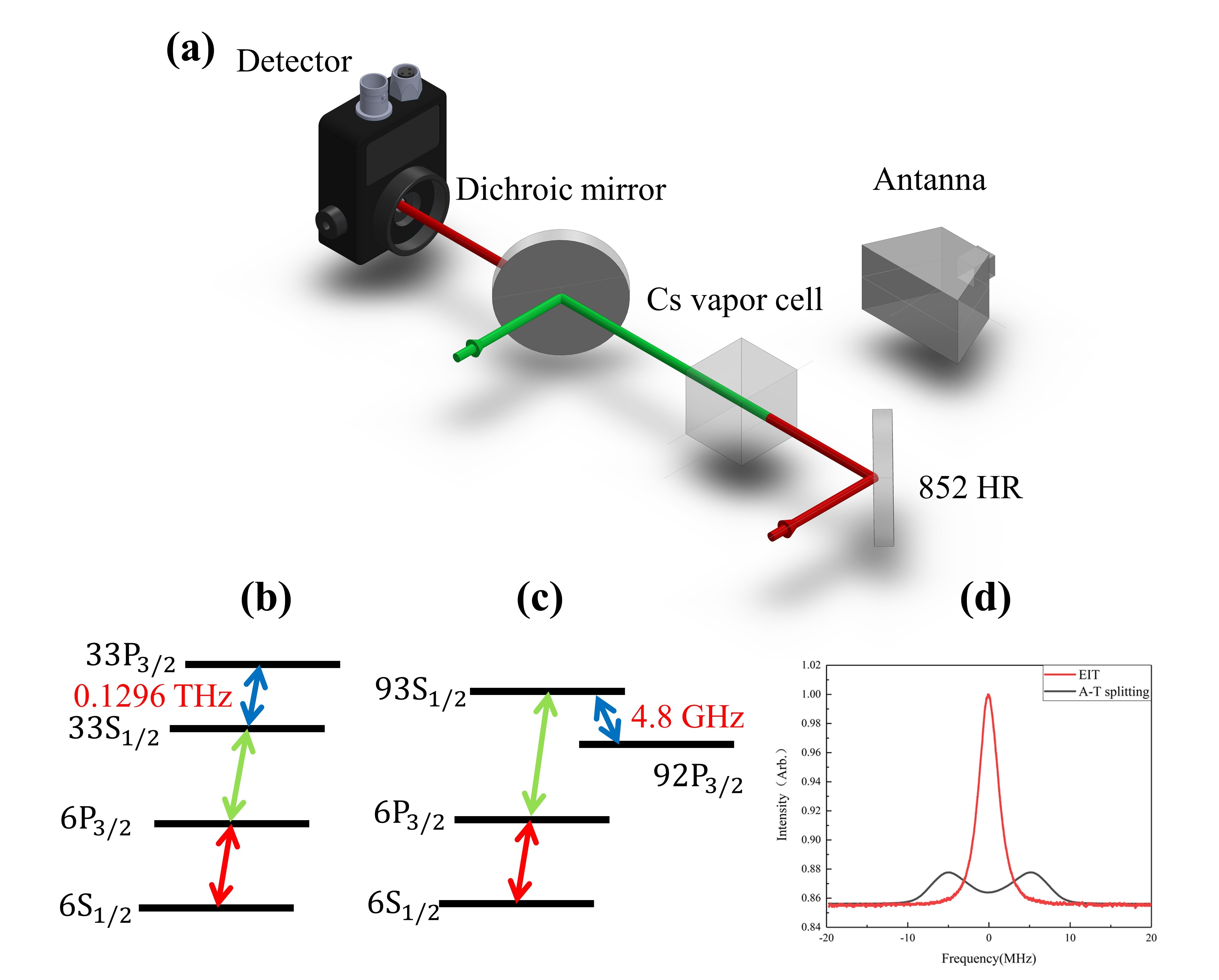}
\caption{\label{fig2}Overall experimental configuration. a). Experimental setup. The red beam represents the 852 nm probe laser, and the green beam represents the 510 nm coupling laser. HR: High Reflection dielectric mirror. b). Energy level of terahertz measurement. c). Energy level of MW measurement. d). An example of EIT spectrum and A-T splitting spectrum, measured in terahertz measurement configuration.}
\end{figure}

We then perform a demonstrative experiment to further illustrate the atomic antenna's isotropic response. In the experiment, the energy differences between dressed states formed by the interaction of the RF field and Rydberg states, i.e., $\Delta_{\rm AT}$, are obtained by measuring the transmission spectrum of the probe laser. Figure \ref{fig2}(a) shows the experimental setup. The linearly polarized probe laser and coupling laser coincide inside a room-temperature cesium vapor cell to prepare the cesium atoms into the Rydberg state by two-step excitation. The probe laser is finally incident into a photodetector so that its transmission spectrum can be recorded when the frequency of the coupling laser is scanned. The probe laser and the coupling laser have opposite propagation directions along X-axis to minimize the effect of Doppler mismatch on the probe laser's transmission spectrum. The linear polarized RF field is irradiated by a horn antenna into the vapor cell, and its incident or polarization direction is adjustable by changing the position and angle of the antenna.

We use two energy level configurations, one enables the measurement of RF fields in the terahertz band, and the other enables the measurement of RF fields in the microwave band, as shown in Fig. \ref{fig2}(b) and \ref{fig2}(c), respectively. In both configurations, we use a probe laser with a wavelength of around 852.35 nm to excite ground state atoms into the excited state of cesium D2 transition ($6S_{1/2},F=4\to6P_{3/2},F=5$). For terahertz (microwave) measurement, a coupling laser with a wavelength of around 511.69 (508.64) nm is used to prepare atoms into $33S_{1/2}$ ($93S_{1/2}$) Rydberg state, and a terahertz (microwave) field with a frequency of around 0.1296 THz (4.8 GHz) couples Rydberg transition of $33S_{1/2}\to33P_{3/2}$ ($93S_{1/2}\to92P_{3/2}$). When measuring the terahertz field, we use a borosilicate glass vapor cell with an inner size of 20 mm cubic, and the thickness of the cell wall is 2 mm. The inner size of the vapor cell is changed to 20*20*80 mm$^3$ when measuring the microwave field to increase the atomic sample's length, which can compensate for the reduction of the spectral brightness caused by the decrease of the coupling laser Rabi frequency for such highly principal quantum number Rydberg state excitation. The laser and energy level configurations in this experimental setup enable a phenomenon of electromagnetically induced transparency (EIT) \cite{Harris1990,Fleischhauer2005}, and the degenerate EIT dark states will split into two states with different energy when Rydberg levels are dressed by RF field coupling, i.e., Autler-Townes  (A-T) splitting phenomenon. Figure \ref{fig2}(d) shows a typical spectrum example of EIT (red) and A-T splitting (black). The frequency interval between A-T peaks reads out $\Delta_{\rm AT}$, and with the known of $\Delta_{\rm AT}$ and $\Delta$, the measurement output of atomic antenna ($E_{\rm MW}$) can be deduced according to Eq. \ref{eq.measurement}.

\begin{figure*}
\includegraphics[width=0.8\textwidth]{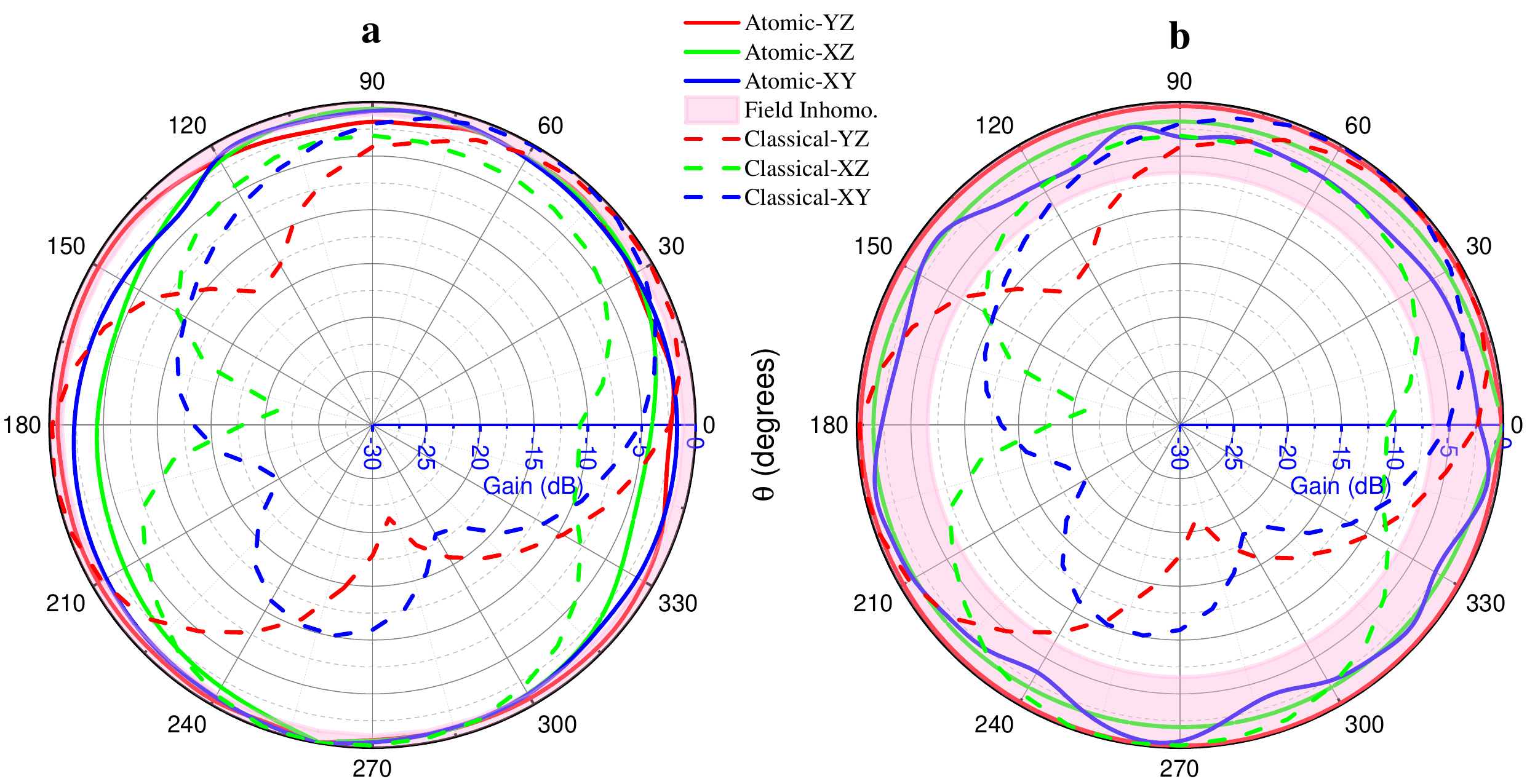}
 \centering
\caption{\label{fig3}The normalized receive-gain patterns of the atomic antenna and classical omnidirectional antenna. a). Gain patterns of atomic antenna (solid line) and classical antenna (dashed line) measured at 4.8 GHz and 4.9 GHz, respectively. The pink shadow shows the simulation measurement uncertainty of the 4.8 GHz field caused by cell geometry's disturbance of field distribution. b). Gain patterns of atomic antenna measured at 0.1296 THz (solid line). The same gain patterns as Fig. \ref{fig3}(a) of classical antenna measured at 4.9 GHz is also shown as a reference. The pink shadow shows the theoretical measurement uncertainty of the 0.1296 THz field caused by cell geometry's disturbance of field distribution.}
\end{figure*}

Whether the atomic antenna has an isotropic response can be verified by measuring normalized gain patterns on three independent orthogonal planes in space, i.e., XY, XZ, and YZ planes. Throughout the overall measurement, we keep the distance of the radiating antenna from the cell, the power injected into the radiating antenna the same, and keep the antenna always pointing to the center of the cell. These ensure that the electric field strength felt by the atoms is always consistent. When measuring the pattern on the XY plane, we move the antenna around the cell on the XY plane for a circle. When measuring the pattern on the XZ plane, for the convenience of the experiment, we point the radiating antenna to the cell along the Y-axis, then rotate the antenna around the Y-axis so that the RF field polarization at the location of the cell rotates around the Y-axis. Since the outcome of atom-field interaction depends only on the polarization direction of the field, this is equivalent to moving and rotating the antenna around the cell in the XZ plane. For the same reason and principle, when measuring the pattern on the YZ plane, instead of moving and rotating the antenna in the YZ plane, we rotate the polarization angle of the linearly polarized excitation laser around the X-axis.

Figure \ref{fig3} shows the experimental measured normalized receive-gain patterns of the atomic antenna. We use the isotropic deviation (gain deviation) \cite{Shah2021} to quantitatively describe the quality of isotropic properties of a receive-sensor, where the isotropic deviation is defined as the difference between the maximum and minimum values of the normalized gain in the same pattern. According to Fig. \ref{fig3}(a), the isotropic deviation of the atomic antenna in XY (blue solid line), XZ (green solid line), and YZ planes (red solid line) for microwave field measurement is about 3.6 dB, 4.5 dB, and 2.5 dB, respectively. The corresponding value for terahertz measurement is about 4.4 dB, 1.4 dB, and 0.28 dB according to Fig. \ref{fig3}(b), respectively. Both isotropic deviations for microwave and terahertz measurements have a minimal value on the YZ plane and are sourced from the measurement uncertainty of $\Delta_{\rm AT}$. The A-T splitting spectrum for microwave measurement has less signal-to-noise ratio than the terahertz measurement spectrum, which leads to much more uncertainty in $\Delta_{\rm AT}$, thus a larger isotropic deviation in YZ plane. The isotropic deviation on other planes is significantly increased relative to the YZ plane, which is mainly due to two reasons: one is that the antenna is artificially moved or rotated in these measurements, which is difficult to guarantee high accuracy, resulting in additional measurement uncertainties; the other is that the field is not uniformly distributed inside the vapor cell. 

\begin{figure*}
\includegraphics[width=\textwidth]{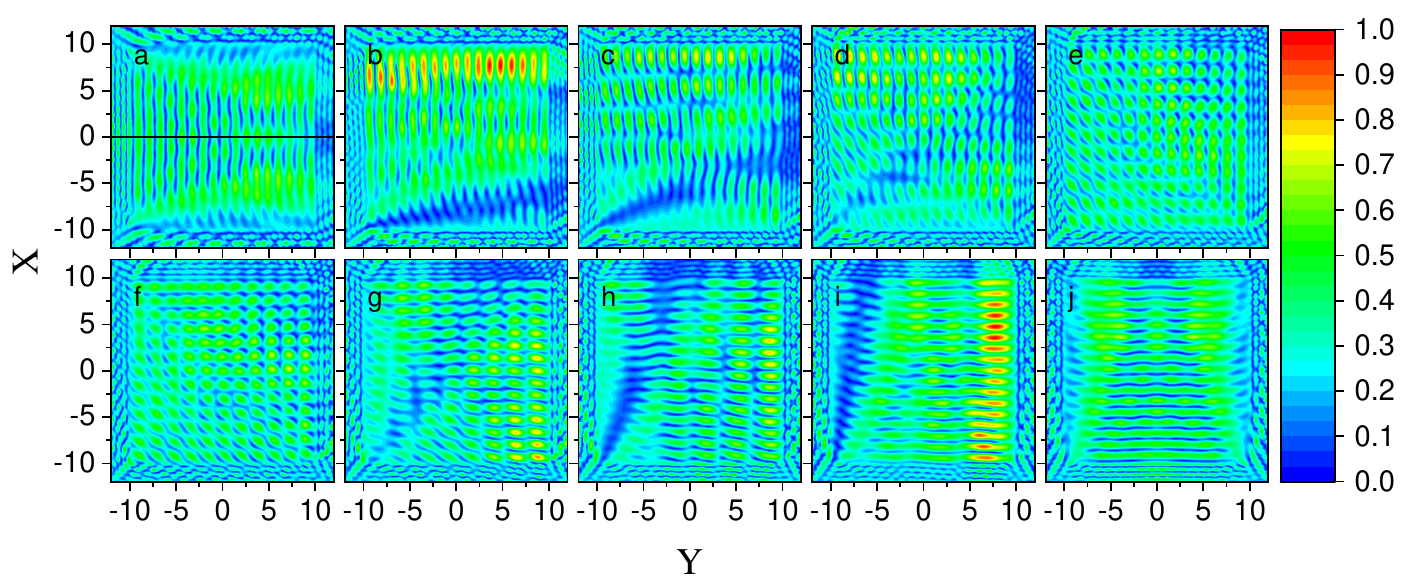}
\caption{\label{fig4} Simulation results of E-field distribution inside a borosilicate glass cell at different incident angles. The field has a frequency of  0.1296 THz. a)-j). The incident angle of the terahertz field is from 0 to 90 degrees, with a stepping of 10 degrees.}
\end{figure*}

To characterize the effect of field inhomogeneity, we used CST software to simulate the distribution of RF fields inside the cell with different field incident angles. The simulation results of field distribution inside the cubic cell of the terahertz field are shown in Fig. \ref{fig4}, which show that the terahertz field forms a standing-wave field distribution inside the 20 mm cubic cell. The distribution pattern of the standing wave field has a pronounced change with the incident direction of the terahertz field. Since the atomic antenna reads out the averaged field along the excitation laser path (black line in Fig. \ref{fig4}(a)), we also averaged the field amplitude along the laser propagation path in the simulation results. After counting this average value at different field incident angles, the deviation of the field strength can be obtained, shown as the pink shadow in Fig. \ref{fig3}(a). Since the simulation and data processing process is the same as that of terahertz, the simulation results of the microwave distribution pattern are not shown, and only the expected deviation caused by the inhomogeneity of field distribution is shown as the pink shadow in Fig. \ref{fig3}(b). 

For the current experiment, the maximum isotropic deviation of the atomic antenna in whole space is within 5 dB, and the isotropic deviation in a single plane can achieve as minimum as 0.28 dB. The largest isotropic deviation is roughly within the error range due to microwave inhomogeneity given by the simulation. At the same time, the minimal isotropic deviation in a single plane proves that with the improvement of experimental accuracy, the measured isotropic deviation can be further reduced and will constantly approach the theoretically predicted value of 0 dB. It shows significant advantages over classical antennas, of which isotropic deviation in whole space is typically larger than 20 dB, as shown in dash lines in Fig. \ref{fig3}. The dash lines in Fig. \ref{fig3} show measurement results of the normalized gain pattern of a classical omnidirectional antenna. This antenna works around  4.9 GHz, and its gain pattern is measured in a microwave anechoic chamber by rotating its direction at a fixed receive position.

\section{Conclusion}
In summary, we have proved that the atomic antenna can achieve perfect isotropic response in principle. This work has conducted a rigorous analysis of the conclusion based on intuitive theory and verified it experimentally. Although the intuitive theory considers RF fields coupling two Rydberg states with the angular momentum of 1/2 and 3/2, the conclusion also applies to other energy-level structures. We verified the theoretical prediction experimentally, achieved an isotropic deviation of less than 5 dB in a demonstration experiment and proved that the experimental isotropic deviation mainly comes from some solvable technical reasons, such as imperfect experimental conditions and the non-uniform field distribution inside the vapor cell. It can further improve to within 0.3 dB by performing the measurement in a better environment and carefully designing the cell geometry. Compared with the isotropic deviation of at least 20 dB of the classical antenna, the isotropic deviation of the atomic antenna realized in this work has an improvement of at least 15 dB. This isotropic atomic antenna is an excellent example of what a tailored quantum sensor can realize but a classical sensor cannot. It will have an essential application in radio frequency electric field metrology.

\nocite{*}

\bibliography{Isotropic_antenna}

\end{document}